\documentclass[preprintnumbers,amsmath,amssymb,floatfix,9pt,prd,onecolumn,superscriptaddress,nofootinbib]{revtex4}
\usepackage{graphicx}
\usepackage{epsfig}
\usepackage{bm}
\usepackage{amsfonts}

\begin{document}

\title{Loop Quantum Corrections to Statefinder Parameters
of Dark Energy}

\author{\textbf{ Mubasher Jamil}} \email{mjamil@camp.nust.edu.pk}
\affiliation{Center for Advanced Mathematics and Physics (CAMP),\\
National University of Sciences and Technology (NUST), H-12,
Islamabad, Pakistan} \affiliation{Eurasian International Center for
Theoretical Physics,  L.N. Gumilyov Eurasian National University,
Astana 010008, Kazakhstan}

\author{\textbf{ D. Momeni}}
\email{d.momeni@yahoo.com } \affiliation{Eurasian International
Center for Theoretical Physics,  L.N. Gumilyov Eurasian National
University, Astana 010008, Kazakhstan}

\author{\textbf{ Ratbay Myrzakulov}}
\email{rmyrzakulov@gmail.com}\affiliation{Eurasian International
Center for Theoretical Physics,  L.N. Gumilyov Eurasian National
University, Astana 010008, Kazakhstan}

\begin{abstract}
\textbf{Abstract:} In this short letter, we presented the explicit forms of  the statefinder
parameters for the Friedmann-Robertson-Walker (FRW) Universe in the
loop quantum cosmology (LQC) for Holographic dark energy and New-Agegraphic dark energy. Numerically we investigated cosmological implications of these parameters for models of DE. 

\end{abstract}
\maketitle

\section{Introduction}

Acceleration expansion is a consequence of observational data
\cite{Perlmutter1, Spergel1}. The role of this epoch is for dark
energy (DE) \cite{Riess1}. A new stringy motivated model of DE is Holographic dark energy(HDE). Also recently it has been proposed  New-agegraphic
dark energy (NADE) from a quantum mechanical point of view.
To classify models of DE Sahni et al \cite{Sahni1}
introduced the statefinder pair
$\left\{r, s\right\}$ given by ,
\begin{equation}\label{state8.1}
r\equiv\frac{\stackrel{...}a}{aH^3},~~~~~~~~~~s\equiv\frac{r-1}{3(q-1/2)},
\end{equation}
As usual $a,H$ denote the scale factor and the Hubble
parameter. Also $q$ is the deceleration parameter is defined by,
\begin{equation}\label{state8.2}
q=-\frac{\stackrel{..}a}{aH^{2}}.
\end{equation}
 In
a model of DE trajectory in $r,s$ plane classifies the model's behavior \cite{Albert1,
Albert2}. \par
Following the modification proposal of gravity it is possible to label the DE to gravity corrections \cite{Nojiri:2010wj}-\cite{Nojiri:2006ri}. Our main goal in this paper is to study this pair in   Loop quantum gravity (LQG) was
introduced \cite{ashtekar} andwas developed \cite{sergei,epjc}. 

The  plan of this paper is as following: In Section 2 we have calculated
the deceleration parameter and statefinder parameters for Loop
quantum gravity. Section 3 deals with Holographic dark energy(HDE)
and in section 4 we do the evaluations for New Agegraphic dark
energy(NADE). In section 5 we provide a detailed and comparative
graphical analysis. Finally the paper ends with a general
discussion in section 6.

\section{Loop Quantum Gravity Model}

Following the idea of loop Quantum Gravity (LQG) , universe evolved as an
outstanding effort to describe the quantum effect of our universe
\cite{Rovelli1, Ashtekar1,Ashtekar2, Bojowald1, Ashtekar3}. Recently it has been reported that loop quantum cosmology is in relation to the teleparallelism mechanism of gravity as a geometrical theory on torsion \cite{sergei}.

Following the cosmological evolution of  LQC  \cite{Wu2} the new
Friedmann equation reads
\cite{Wu1,Chen1,Fu1}
\begin{equation}\label{state8.3}
H^2=\frac{\kappa^2}{3}\rho\left(1-\frac{\rho}{\rho_{1}}\right).
\end{equation}
Here $\kappa^2=8\pi G$ and
$\rho_{1}=\sqrt{3}\pi^{2}\gamma^{3}G^{2}\hbar$ defines the "`critical loop
quantum density"' and $\gamma$ is the dimensionless Barbero-Immirzi
parameter. $\rho=\rho_{m}+\rho_{D}$~ represents the total cosmic
energy density, which is a sum of energy density of DM
(${\rho}_{m}$) and the energy density of DE ($\rho_{D}$).

The conservation equations  are
\begin{equation}\label{state8.4}
\dot{\rho}_{D}+3H\left(1+\omega_{D}\right)\rho_{D}=0,
\end{equation}
\begin{equation}\label{state8.5}
\dot{\rho}_m+3H\rho_m=0.
\end{equation}
Using (\ref{state8.3}) and the conservation
identities, we can obtain the modified geometric Raychaudhuri equation
\begin{equation}\label{state8.6}
\dot{H}=-\frac{\kappa^{2}}{2}\left(\rho+p\right)\left(1-2\frac{\rho}{\rho_{1}}\right).
\end{equation}
Let the dark energy obey the equation of state~
$p_{D}=\omega_D\rho_{D}$. Using this expression for $p_{D}$ and
using the expression $\rho_{T}=\rho=\rho_{D}+\rho_m$ for total
cosmic energy density $\rho$ we get,
\begin{equation}\label{state8.7}
\dot{H}=-\frac{\kappa^{2}}{2}\left(\rho_{D}+\rho_m+\omega
\rho_{D}\right)\left(1-\frac{2\left(\rho_{D}+\rho_m\right)}{\rho_{1}}\right).
\end{equation}

We define two dimensionless density parameters as follows,
\begin{equation}\label{state8.8}
\Omega_{D}=\frac{\kappa^{2}\rho_{D}}{3H^{2}},~~~~~~~~~\Omega_{m}=\frac{\kappa^{2}\rho_{m}}{3H^{2}}.
\end{equation}
The deceleration parameter in terms of these dimensionless density
parameters is given by,
\begin{equation}\label{state8.9}
q=-1+\frac{3}{2}\left(2-\Omega_{m}-\Omega_{D}\right)\left(1+\frac{\omega_{D}\Omega_{D}}{\Omega_{m}+\Omega_{D}}\right).
\end{equation}
 From the expression of $r$, we have a
relation between $r$ and deceleration parameter $q$ as,
\begin{equation}\label{state8.10}
r=2q^{2}+q-\frac{\dot{q}}{H}.
\end{equation}
The time derivatives of the dimensionless density parameters
$\Omega_{m}$ and $\Omega_{D}$ are given as,
\begin{equation}\label{state8.11}
\dot{\Omega}_{m}=\Omega_{m}H\left(2q-1\right).
\end{equation}
\begin{equation}\label{state8.12}
\dot{\Omega}_{D}=\Omega_{D}H\left(2q-3\omega_{D}-1\right).
\end{equation}
Using equations (\ref{state8.8}) to (\ref{state8.12}) we get,

$$r=2\left[-1+\frac{3}{2}\left(2-\Omega_{m}-\Omega_{D}\right)\left(1+\frac{\omega_{D}\Omega_{D}}{\Omega_{m}+\Omega_{D}}\right)\right]^{2}-1+\frac{3}{2}\left(2-\Omega_{m}-\Omega_{D}\right)\left(1+\frac{\omega_{D}\Omega_{D}}{\Omega_{m}+\Omega_{D}}\right)$$
$$-\frac{1}{H}\left[\frac{3}{2}\left\{-\Omega_{m}H\left(-3+3\left(2-\Omega_{m}-\Omega_{D}\right)\left(1+\frac{\omega_{D}\Omega_{D}}{\Omega_{m}+\Omega_{D}}\right)\right)-\Omega_{D}H\left(-3-3\Omega_{D}+3\left(2-\Omega_{m}-\Omega_{D}\right)\right.\right.\right.$$
$$\left.\left.\left.\left(1+\frac{\omega_{D}\Omega_{D}}{\Omega_{m}+\Omega_{D}}\right)\right)\right\}\right]\left(1+\frac{\omega_{D}\Omega_{D}}{\Omega_{m}+\Omega_{D}}\right)+\frac{3}{2}\left[2-\Omega_{m}-\Omega_{D}\left\{\frac{\dot{\omega_{D}}\Omega_{D}}{\Omega_{m}+\Omega_{D}}+\frac{\omega_{D}\Omega_{D}H}{\Omega_{m}+\Omega_{D}}\left(-3-3\Omega_{D}\right.\right.\right.$$
$$\left.\left.\left.+3\left(2-\Omega_{m}-\Omega_{D}\right)\left(1+\frac{\omega_{D}\Omega_{D}}{\Omega_{m}+\Omega_{D}}\right)\right)-\frac{1}{\left(\Omega_{m}+\Omega_{D}\right)^{2}}\left(\omega_{D}\Omega_{D}\left(\Omega_{m}H\left(-3+3\left(2-\Omega_{m}-\Omega_{D}\right)\right.\right.\right.\right.\right.$$
\begin{equation}\label{state8.13}
\left.\left.\left.\left.\left.\left(1+\frac{\omega_{D}\Omega_{D}}{\Omega_{m}+\Omega_{D}}\right)\right)+\Omega_{D}H\left(-3-3\Omega_{D}+3\left(2-\Omega_{m}-\Omega_{D}\left(1+\frac{\omega_{D}\Omega_{D}}{\Omega_{m}+\Omega_{D}}\right)\right)\right)\right)\right)\right\}\right].
\end{equation}

$$s=\left[\frac{1}{-3+\frac{9}{2}\left(2-\Omega_{m}-\Omega_{D}\left(1+\frac{\omega_{D}\Omega_{D}}{\Omega_{m}+\Omega_{D}}\right)-\frac{3}{2}\Omega_{m}\left(1-\frac{3H^{2}\left(\Omega_{m}+\Omega_{D}\right)}{\rho_{1}}\right)-\frac{3}{2}\Omega_{m}\left(1-\frac{3H^{2}\left(\Omega_{m}+\Omega_{D}\right)}{\rho_{1}}\right)\right)}\right]$$
$$\left[2\left\{-1+\frac{3}{2}\left(2-\Omega_{m}-\Omega_{D}\right)\left(1+\frac{\omega_{D}\Omega_{D}}{\Omega_{m}+\Omega_{D}}\right)\right\}^{2}-1+\frac{3}{2}\left(2-\Omega_{m}-\Omega_{D}\right)\left(1+\frac{\omega_{D}\Omega_{D}}{\Omega_{m}+\Omega_{D}}\right)\right.$$
$$\left.-\frac{1}{H}\left[\frac{3}{2}\left\{-\Omega_{m}H\left(-3+3\left(2-\Omega_{m}-\Omega_{D}\right)\left(1+\frac{\omega_{D}\Omega_{D}}{\Omega_{m}+\Omega_{D}}\right)\right)-\Omega_{D}H\left(-3-3\Omega_{D}+3\left(2-\Omega_{m}-\Omega_{D}\right)\right.\right.\right.\right.$$
$$\left.\left.\left.\left.\left(1+\frac{\omega_{D}\Omega_{D}}{\Omega_{m}+\Omega_{D}}\right)\right)\right\}\right]\left(1+\frac{\omega_{D}\Omega_{D}}{\Omega_{m}+\Omega_{D}}\right)+\frac{3}{2}\left[2-\Omega_{m}-\Omega_{D}\left\{\frac{\dot{\omega_{D}}\Omega_{D}}{\Omega_{m}+\Omega_{D}}+\frac{\omega_{D}\Omega_{D}H}{\Omega_{m}+\Omega_{D}}\left(-3-3\Omega_{D}\right.\right.\right.\right.$$
$$\left.\left.\left.\left.+3\left(2-\Omega_{m}-\Omega_{D}\right)\left(1+\frac{\omega_{D}\Omega_{D}}{\Omega_{m}+\Omega_{D}}\right)\right)-\frac{1}{\left(\Omega_{m}+\Omega_{D}\right)^{2}}\left(\omega_{D}\Omega_{D}\left(\Omega_{m}H\left(-3+3\left(2-\Omega_{m}-\Omega_{D}\right)\right.\right.\right.\right.\right.\right.$$
$$\left.\left.\left.\left.\left.\left.\left(1+\frac{\omega_{D}\Omega_{D}}{\Omega_{m}+\Omega_{D}}\right)\right)+\Omega_{D}H\left(-3-3\Omega_{D}+3\left(2-\Omega_{m}-\Omega_{D}\left(1+\frac{\omega_{D}\Omega_{D}}{\Omega_{m}+\Omega_{D}}\right)\right)\right)\right)\right)\right\}\right]\right.$$
\begin{equation}\label{state8.14}
\left.-\Omega_{m}\left(1-\frac{3H^{2}\left(\Omega_{m}+\Omega_{D}\right)}{\rho_{1}}\right)-\Omega_{D}\left(1-\frac{3H^{2}\left(\Omega_{m}+\Omega_{D}\right)}{\rho_{1}}\right)\right].
\end{equation}

\section{Holographic Dark Energy}

For HDE  \cite{Susskind1, Hooft1}following the naive idea of Cohen et al \cite{Cohen1} 
we write the following equation for DE energy density:
\begin{equation}\label{state8.16}
\rho_{vac}\sim\Lambda_{UV}^{4}\sim M_{p}^{2}L^{-2}.
\end{equation}
or equivalently \cite{Li1}
\begin{equation}\label{state8.17}
\rho_{H}=3c^{2}M_{p}^{2}L^{-2},
\end{equation}
where $c\sin O(1)$ is the holographic parameter  and L is the
IR-cutoff of the universe. with future event horizon of the universe as the IR cutoff,
we have:
\begin{equation}\label{state8.18}
R_{h}=a\int_{t}^{\infty}\frac{dt}{a}=a\int_{a}^{\infty}\frac{da'}{Ha'^{2}}.
\end{equation}
Using equation (\ref{state8.17}) and conservation equation
(\ref{state8.4}) the equation of state(EoS) parameter for
holographic dark energy is given by,
\begin{equation}\label{state8.19}
\Omega_{D}=\frac{1}{3}\left(-1-\frac{2}{c}\sqrt{\Omega_{D}}\right),
\end{equation}
where $\Omega_{D}$ is the dimensionless density parameter of
holographic dark energy(HDE) and the time derivative of the
dimensionless density parameter for HDE is given by,
\begin{equation}\label{state8.20}
\dot{\Omega_{D}}=\Omega_{D}H\left(1-\Omega_{D}\right)\left(1+\frac{2}{c}\sqrt{\Omega_{D}}\right).
\end{equation}

Now the expression for the state parameter $r$ is given by,

$$r=\frac{1}{2c^{2}\left(\Omega_{m}+\Omega_{D}\right)^{3}}\left[-6\Omega_{m}^{2}c\left(\frac{19}{3}-\frac{43}{6}\Omega_{m}+\Omega_{m}^{2}\right)\Omega_{D}^{\frac{3}{2}}+10\Omega_{m}^{6}-9c\Omega_{D}^{\frac{11}{2}}+\left(30\Omega_{m}-26+2c^{2}\right)\Omega_{D}^{5}\right.$$
$$\left.\left(25c-33c\Omega_{m}\right)\Omega_{D}^{\frac{9}{2}}+\left(20+30\Omega_{m}^{2}-8c^{2}+12\Omega_{m}c^{2}-74\Omega_{m}\right)\Omega_{D}^{4}-45\left(-\frac{77}{45}\Omega_{m}+\frac{2}{5}+\Omega_{m}^{2}\right)c\Omega_{D}^{\frac{7}{2}}\right.$$
$$\left.+\left(10\Omega_{m}^{3}+6c^{2}-42\Omega_{m}c^{2}-54\Omega_{m}^{2}+24c^{2}\Omega_{m}^{2}+48\Omega_{m}\right)\Omega_{D}^{3}-27\left(\frac{40}{27}+\Omega_{m}^{2}-\frac{95}{27}\Omega_{m}\right)\Omega_{m}c\Omega_{D}^{\frac{5}{2}}\right.$$
$$\left.+\left\{\left(20c^{2}-6\right)\Omega_{m}^{3}+\left(-74c^{2}+12\right)\Omega_{m}^{2}+28\Omega_{m}c^{2}\right\}\Omega_{D}^{2}+6\Omega_{m}^{2}\left(\Omega_{m}^{2}-\frac{29}{3}\Omega_{m}+\frac{19}{3}\right)c^{2}\Omega_{D}\right.$$
\begin{equation}\label{state8.21}
\left.-18\Omega_{m}^{3}\left(\Omega_{m}-\frac{10}{9}\right)c^{2}\right].
\end{equation}

The expression for the state parameter $s$ is given by,

$$s=\left[\left\{-\frac{1}{2}c\left(\Omega_{m}+\Omega_{D}\right)\Omega_{m}\left(1-\frac{3H^{2}\left(\Omega_{m}+\Omega_{D}\right)}{\rho_{1}}\right)-\frac{1}{2}c\left(\Omega_{m}+\Omega_{D}\right)\Omega_{D}\left(1-\frac{3H^{2}\left(\Omega_{m}+\Omega_{D}\right)}{\rho_{1}}\right)\right.\right.$$
$$\left.\left.+\Omega_{D}^{\frac{5}{2}}-c\Omega_{D}^{2}+\left(-2+\Omega_{m}\right)\Omega_{D}^{\frac{3}{2}}+\left(-1-\frac{5}{2}\Omega_{m}\right)c\Omega_{D}-\frac{3}{2}\Omega_{m}c\left(-\frac{4}{3}+\Omega_{m}\right)\right\}c\left(\Omega_{m}+\Omega_{D}\right)^{2}\right]^{-1}$$
$$\frac{1}{6}\left[-2c^{2}\left(\Omega_{m}+\Omega_{D}\right)^{3}\Omega_{m}\left(1-\frac{3H^{2}\left(\Omega_{m}+\Omega_{D}\right)}{\rho_{1}}\right)-2c^{2}\left(\Omega_{m}+\Omega_{D}\right)^{3}\Omega_{D}\left(1-\frac{3H^{2}\left(\Omega_{m}+\Omega_{D}\right)}{\rho_{1}}\right)\right.$$
$$\left.+\left\{\left(10+\Omega_{m}\right)a-6\Omega_{m}^{2}+42\Omega_{m}-48\right\}c\Omega_{m}^{2}\Omega_{D}^{\frac{3}{2}}+\left\{6\Omega_{D}^{6}-c\Omega_{D}^{\frac{11}{2}}+\left(18\Omega_{m}-2c^{2}-10\right)\Omega_{D}^{5}\right.\right.$$
$$\left.\left.+c\left(-1-3\Omega_{m}\right)\Omega_{D}^{\frac{9}{2}}+\left(\left(2-6\Omega_{m}\right)c^{2}-34\Omega_{m}+4+18\Omega_{m}^{2}\right)\Omega_{D}^{4}-3\left(\Omega_{m}^{2}+3\Omega_{m}-\frac{2}{3}\right)c\Omega_{D}^{\frac{7}{2}}\right.\right.$$
$$\left.\left.-6\Omega_{m}\left(\left(\Omega_{m}-\frac{2}{3}\right)c^{2}-\Omega_{m}^{2}-\frac{8}{3}+5\Omega_{m}\right)\Omega_{D}^{3}-c\Omega_{m}\left(-12+\Omega_{m}^{2}+7\Omega_{m}\right)\Omega_{D}^{\frac{5}{2}}-2\left(\left(-1-2\Omega_{m}+\Omega_{m}^{2}\right)c^{2}\right.\right.\right.$$
$$\left.\left.\left.-6\Omega_{m}+3\Omega_{m}^{2}\right)\Omega_{m}\Omega_{D}^{2}+2c^{2}\Omega_{m}^{2}\left(\Omega_{m}+1\right)\Omega_{D}\right\}a+4\Omega_{D}^{6}-8c\Omega_{D}^{\frac{11}{2}}+\left(-16+12\Omega_{m}+4c^{2}\right)\Omega_{D}^{5}\right.$$
$$\left.+\left(-30\Omega_{m}+26\right)c\Omega_{D}^{\frac{9}{2}}+\left(\left(18\Omega_{m}-10\right)c^{2}+12\Omega_{m}^{2}+16-40\Omega_{m}\right)\Omega_{D}^{4}+\left(86\Omega_{m}-42\Omega_{m}^{2}-20\right)c\Omega_{D}^{\frac{7}{2}}\right.$$
$$\left.+\left(\left(-46\Omega_{m}+30\Omega_{m}^{2}+6\right)c^{2}+32\Omega_{m}+4\Omega_{m}^{3}-24\Omega_{m}^{2}\right)\Omega_{D}^{3}-26\Omega_{m}\left(\Omega_{m}^{2}-\frac{51}{13}\Omega_{m}+2\right)c\Omega_{D}^{\frac{5}{2}}\right.$$
\begin{equation}\label{state8.22}
\left.+22\Omega_{m}\left(-\frac{39}{11}\Omega_{m}+\frac{13}{11}+\Omega_{m}^{2}\right)c^{2}
\Omega_{D}^{2}+6c^{2}\Omega_{m}^{2}\left(-10\Omega_{m}+6+\Omega_{m}^{2}\right)\Omega_{D}-18\Omega_{m}^
{3}\left(\Omega_{m}-\frac{10}{9}\right)c^{2}\right].
\end{equation}

\begin{figure*}[thbp]
\begin{tabular}{rl}
\includegraphics[width=7cm]{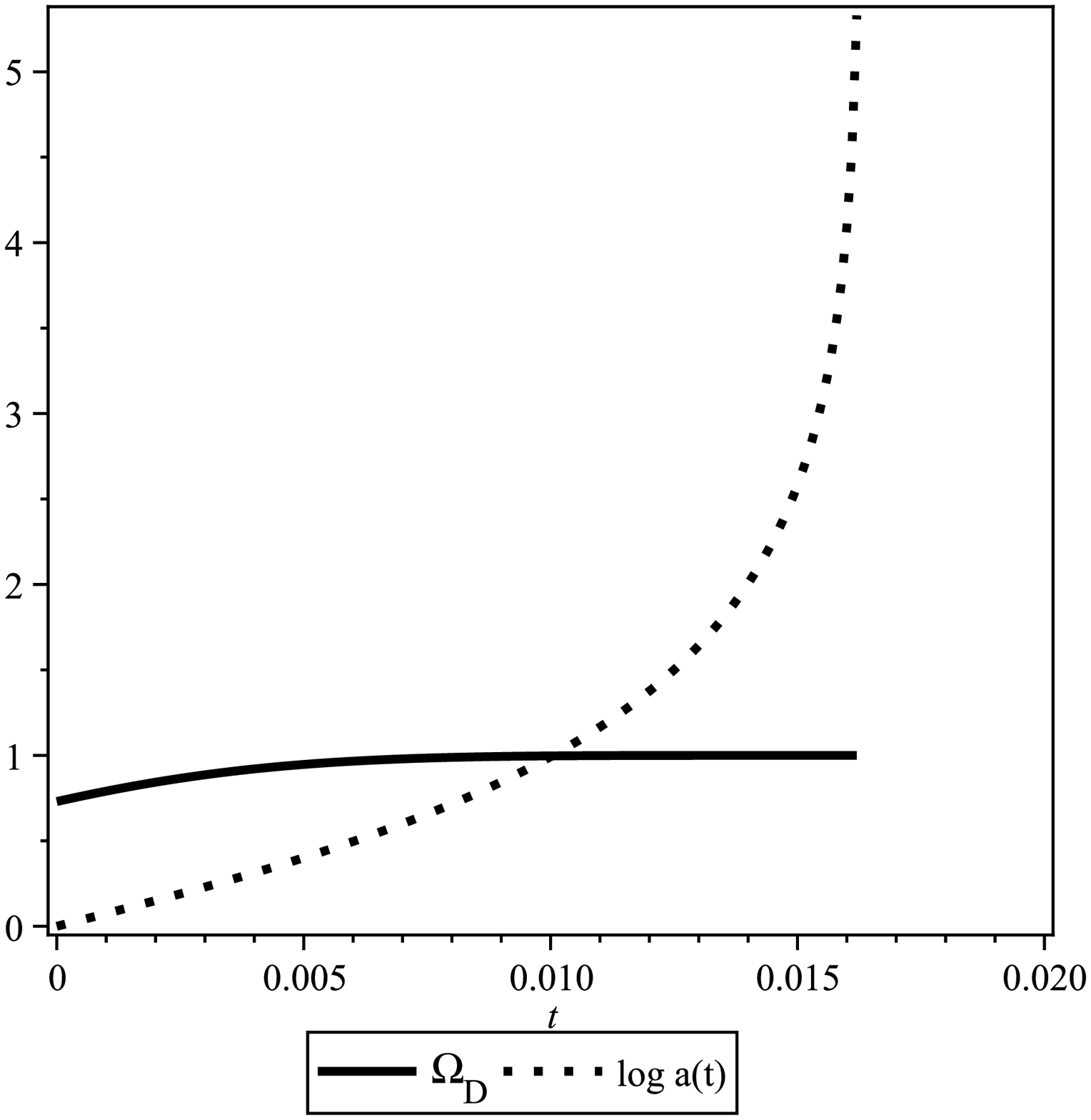}&
\includegraphics[width=7cm]{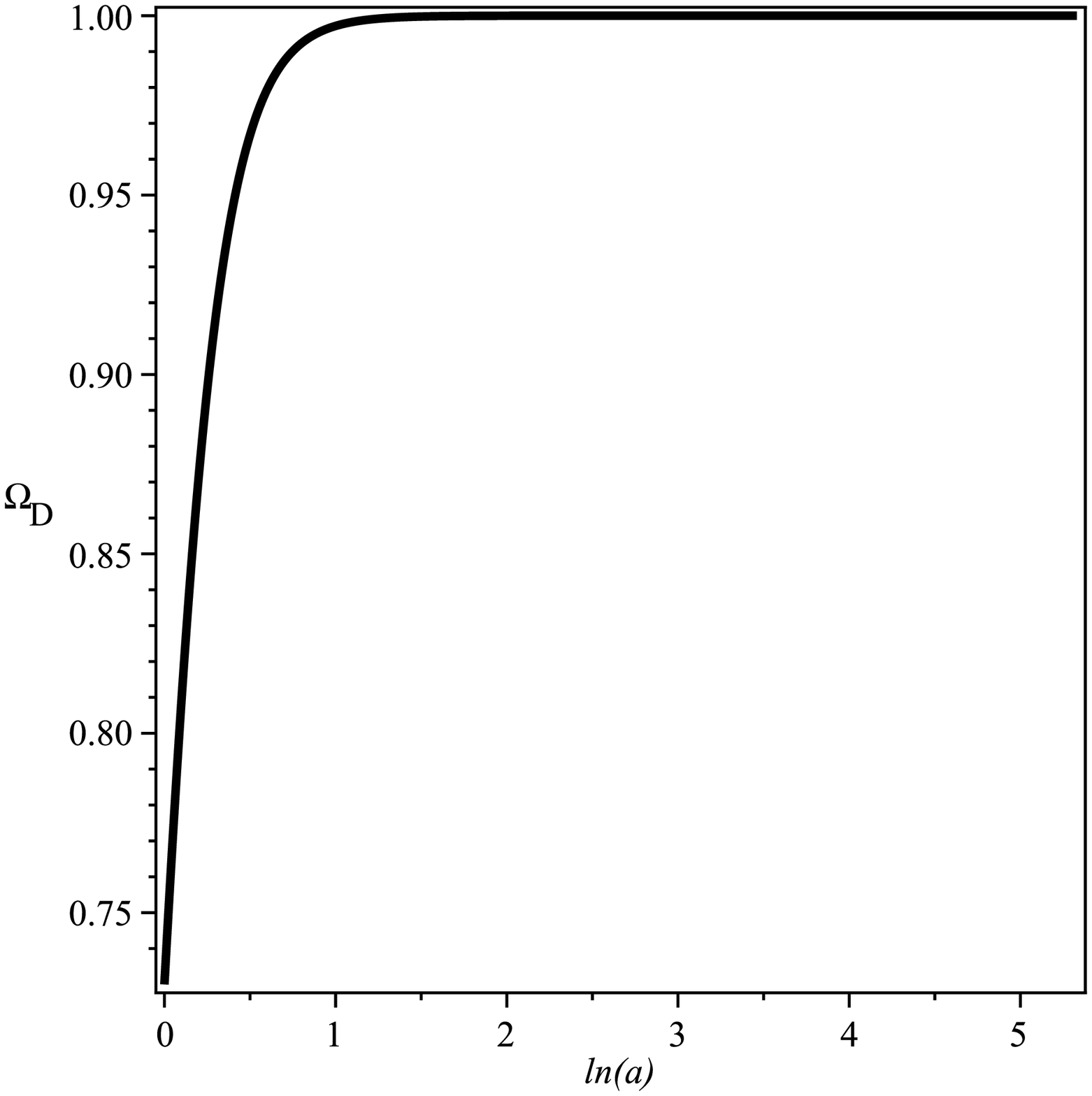} \\
\includegraphics[width=7cm]{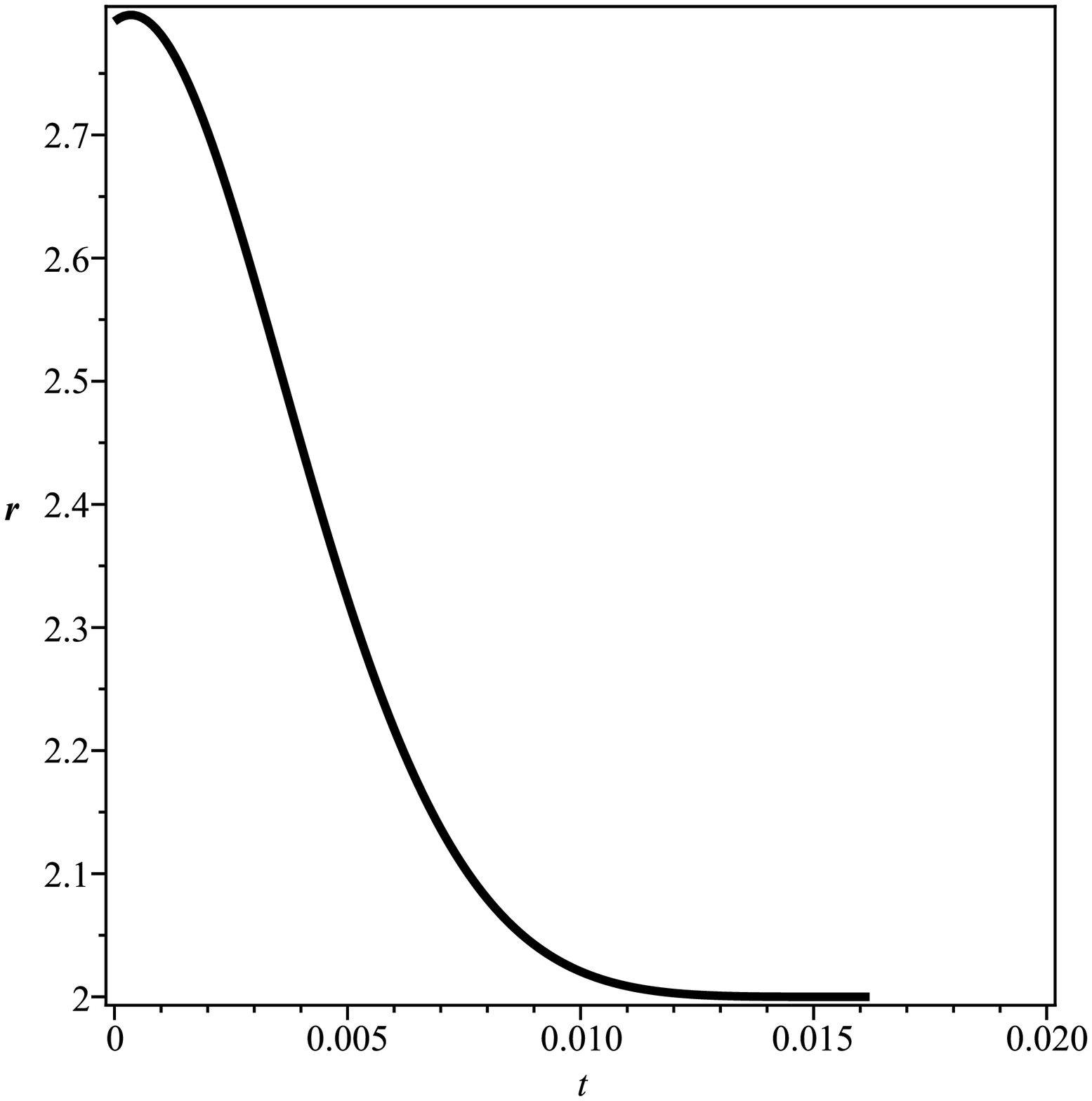}&
\includegraphics[width=7cm]{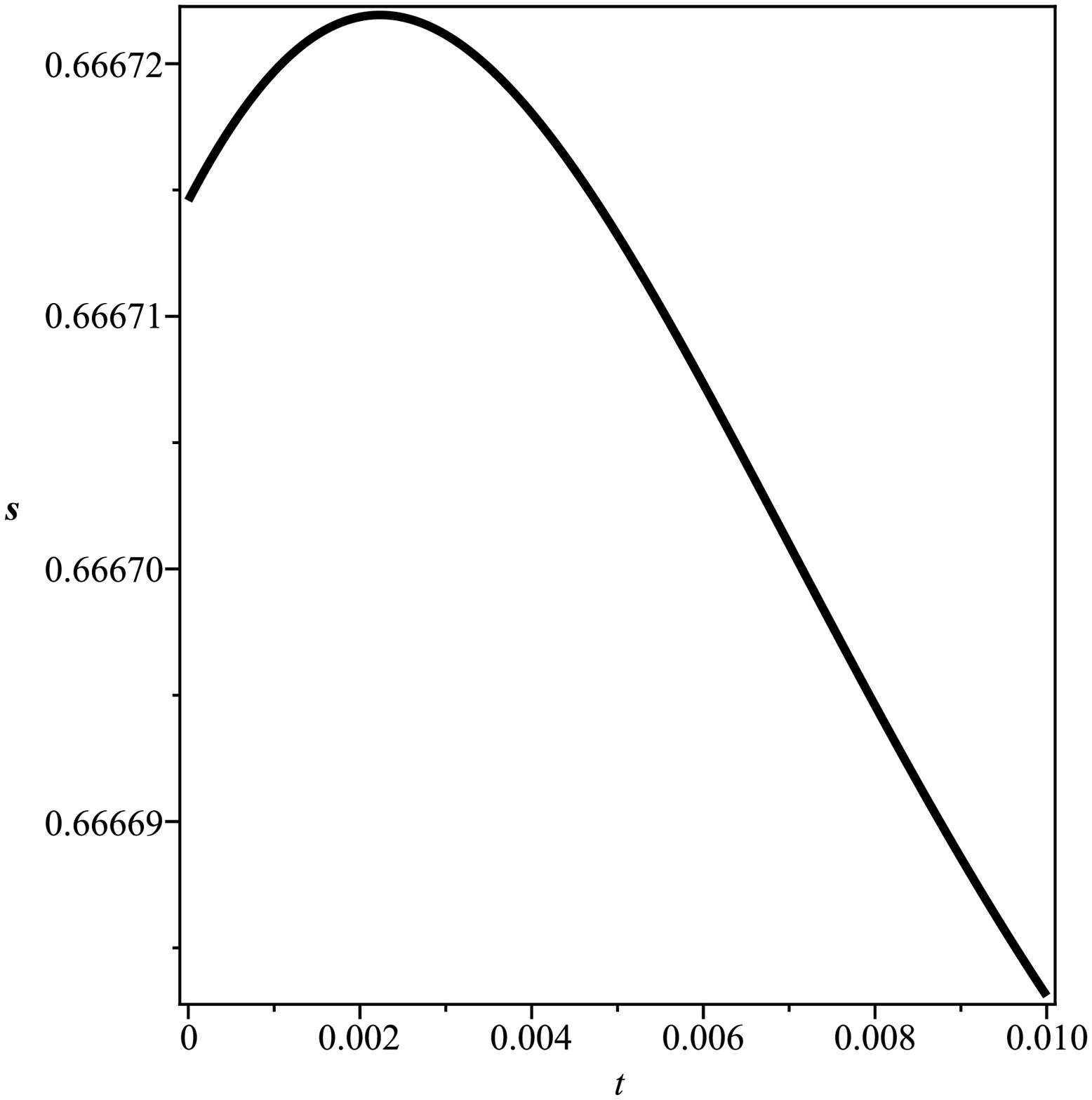} \\
\end{tabular}
\caption{(\textit{Top Left}) Time evolution of HDE density parameter
and logarithmic scale factor.    (\textit{Top Right}) Evolutionary
behavior of HDE density parameter versus logarithmic scale factor.
(\textit{Bottom Left}) Time evolution of statefinder parameter $r$.
(\textit{Bottom Right}) Time evolution of statefinder parameter $s$.
While plotting, we adopted the following fixing of free parameters:
$a(0)=1$, $\dot{a}(0)=H_0=74$, $\Omega_D(0)=0.73$, $c=0.5$,
$\rho_1=0.187$.  }
\end{figure*}

\section{New Agegraphic Dark Energy}

Now for 
Agegraphic dark energy \cite{Cai1,Wei2, Zhang1,Sasakura1, Cai1, Wei2, Wei3,Karolyhazy1,
Karolyhazy2}we write \cite{Maziashvili1, Maziashvili2},
\begin{equation}\label{state8.23}
\rho_{\Lambda}\sim\frac{1}{t_{2}t_{p}^{2}}\sim\frac{M_{p}^{2}}{t^{2}},
\end{equation}
Now $M_{p}$ is the reduced planck mass.
So the dark energy density $\rho_{A}$ of NADE model
\cite{Karolyhazy1, Karolyhazy2} has the form,
\begin{equation}\label{state8.24}
\rho_{A}=3n^{2}M_{p}^{2}\eta^{-2},
\end{equation}
where $3n^{2}\sim\mathcal{O}(1)$  and $\eta$ is a conformal
time .

Combining equations (\ref{state8.22}), (\ref{state8.23}) and
conservation equation (\ref{state8.2}) the NADE equation of
state(EoS) parameter is obtained as,
\begin{equation}\label{state8.26}
\Omega_{D}=-1+\frac{2}{3na}\sqrt{\Omega_{D}}.
\end{equation}
The time derivative of the dimensionless density parameter of NADE
is given by,
\begin{equation}\label{state8.27}
\dot{\Omega_{D}}=\Omega_{D}H\left(1-\Omega_{D}\right)\left(3-\frac{2}{na}\sqrt{\Omega_{D}}\right).
\end{equation}

The expression for the state parameter $r$ is given by,
$$r=\frac{1}{2n^{2}a^{2}\left(\Omega_{m}+\Omega_{D}\right)^{3}}\left[6na\left(3-\frac{11}{2}\Omega_{m}+\Omega_{m}^{2}\right)\Omega_{m}^{2}\Omega_{D}^{\frac{3}{2}}+10\Omega_{D}^{6}-9na\Omega_{D}^{\frac{11}{2}}+\left(30\Omega_{m}-26\right)\Omega_{D}^{5}\right.$$
$$\left.-21na\left(-1+\Omega_{m}\right)\Omega_{D}^{\frac{9}{2}}+\left(20+30\Omega_{m}^{2}-74\Omega_{m}\right)\Omega_{D}^{4}-9\left(\Omega_{m}-6\right)n\left(\Omega_{m}-\frac{1}{3}\right)a\Omega_{D}^{\frac{7}{2}}+\left(10\Omega_{m}^{3}-54\Omega_{m}^{2}\right.\right.$$
$$\left.\left.+\left(48-18n^{2}a^{2}\right)\Omega_{m}+2n^{2}a^{2}\right)\Omega_{D}^{3}+9n\left(-\frac{16}{3}+\frac{1}{3}\Omega_{m}+\Omega_{m}^{2}\right)a\Omega_{m}\Omega_{D}^{\frac{5}{2}}+\left(-6\Omega_{m}^{3}+\left(-18n^{2}a^{2}+12\right)\Omega_{m}^{2}\right.\right.$$
\begin{equation}\label{state8.28}
\left.\left.+24n^{2}a^{2}\Omega_{m}\right)\Omega_{D}^{2}-18n^{2}
\left(\Omega_{m}-\frac{1}{3}\right)a^{2}\Omega_{m}^{2}\Omega_{D}-18n^{2}a^{2}
\Omega_{m}^{3}\left(\Omega_{m}-\frac{10}{9}\right)\right].
\end{equation}

The expression for the state parameter $s$ is given by,

$$s=\left[\left(\Omega_{m}+\Omega_{D}\right)^{2}\left\{\frac{1}{2}na\left(\Omega_{m}+\Omega_{D}\right)\Omega_{m}\left(1-\frac{3H^{2}\left(\Omega_{m}+\Omega_{D}\right)}{\rho_{1}}\right)+\frac{1}{2}na\left(\Omega_{m}+\Omega_{D}\right)\Omega_{D}\right.\right.$$
$$\left.\left.\left(1-\frac{3H^{2}\left(\Omega_{m}+\Omega_{D}\right)}{\rho_{1}}\right)+\left(-2+\Omega_{m}\right)\Omega_{D}^{\frac{3}{2}}+\Omega_{D}^{\frac{5}{2}}+\frac{3}{2}na\left(\frac{2}{3}+\Omega_{m}\right)\Omega_{D}+\frac{3}{2}na\left(-\frac{4}{3}+\Omega_{m}\right)\Omega_{m}\right\}na\right]^{-1}$$
$$\frac{1}{6}\left[2n^{2}a^{2}\left(\Omega_{m}+\Omega_{D}\right)^{3}\Omega_{m}\left(1-\frac{3H^{2}\left(\Omega_{m}+\Omega_{D}\right)}{\rho_{1}}\right)+2n^{2}a^{2}\left(\Omega_{m}+\Omega_{D}\right)^{3}\Omega_{D}\left(1-\frac{3H^{2}\left(\Omega_{m}+\Omega_{D}\right)}{\rho_{1}}\right)\right.$$
$$\left.-6na\left(3-\frac{11}{2}\Omega_{m}+\Omega_{m}^{2}\right)\Omega_{m}^{2}\Omega_{D}^{\frac{3}{2}}-10\Omega_{D}^{6}+9na\Omega_{D}^{\frac{11}{2}}+\left(26-30\Omega_{m}\right)\Omega_{D}^{5}+21na\left(-1+\Omega_{m}\right)\Omega_{D}^{\frac{9}{2}}\right.$$
$$\left.+\left(-30\Omega_{m}^{2}-20+74\Omega_{m}\right)\Omega_{D}^{4}+9\left(\Omega_{m}-6\right)n\left(\Omega_{m}-\frac{1}{3}\right)a\Omega_{D}^{\frac{7}{2}}+\left(-10\Omega_{m}^{3}+54\Omega_{m}^{2}+\left(-48+18n^{2}a^{2}\right)\Omega_{m}\right.\right.$$
$$\left.\left.-2n^{2}a^{2}\right)\Omega_{D}^{3}-9n\left(-\frac{16}{3}+\frac{1}{3}\Omega_{m}+\Omega_{m}^{2}\right)a\Omega_{m}\Omega_{D}^{\frac{5}{2}}+\left(6\Omega_{m}^{3}+\left(-12+18n^{2}a^{2}\right)\Omega_{m}^{2}-24n^{2}a^{2}\Omega_{m}\right)\Omega_{D}^{2}\right.$$
\begin{equation}\label{state8.29}
\left.+18n^{2}\left(\Omega_{m}-\frac{1}{3}\right)a^{2}\Omega_{m}^{2}\Omega_{D}
+18n^{2}a^{2}\Omega_{m}^{3}\left(\Omega_{m}-\frac{10}{9}\right)\right].
\end{equation}

\begin{figure*}[thbp]
\begin{tabular}{rl}
\includegraphics[width=7cm]{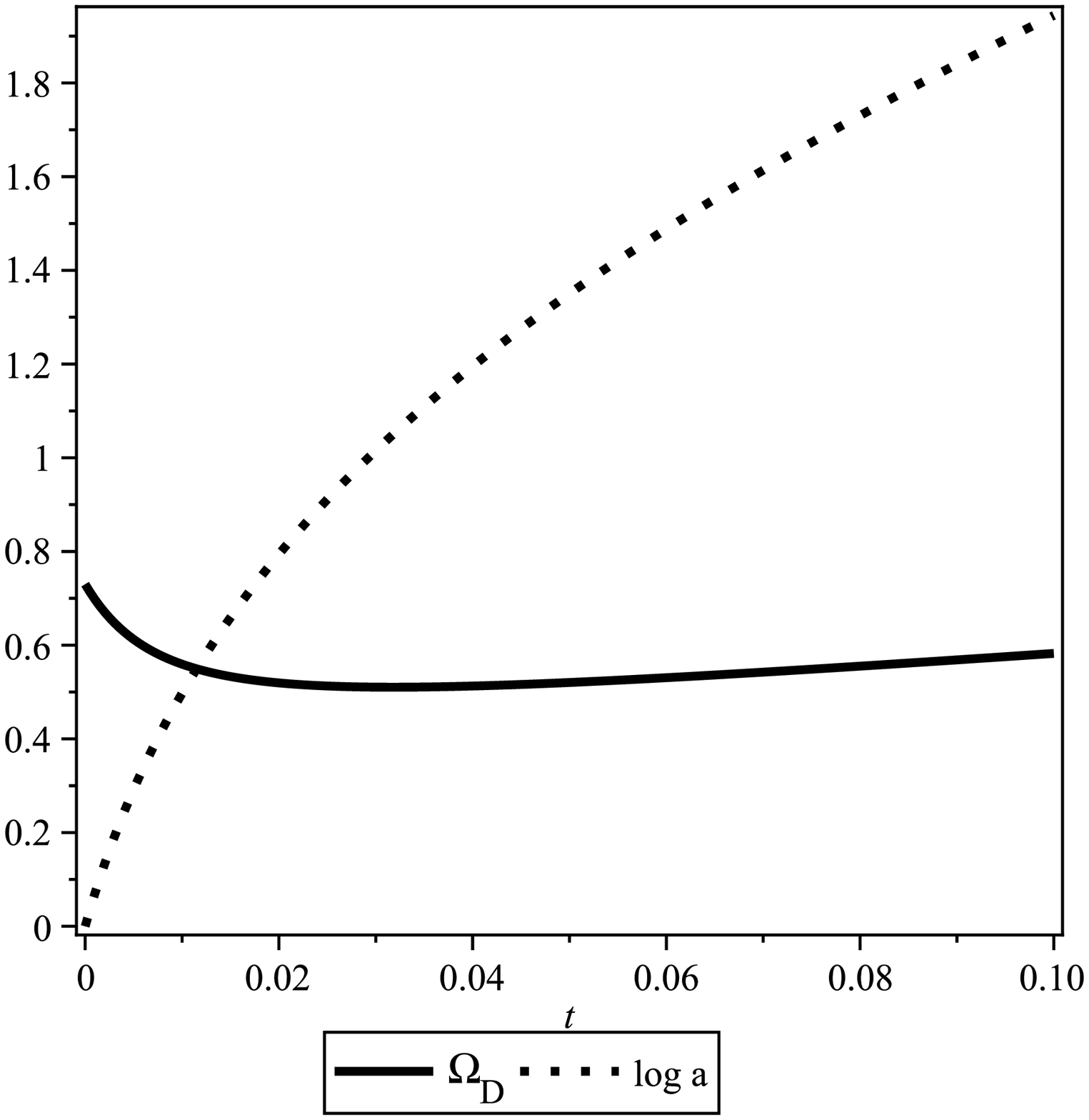}&
\includegraphics[width=7cm]{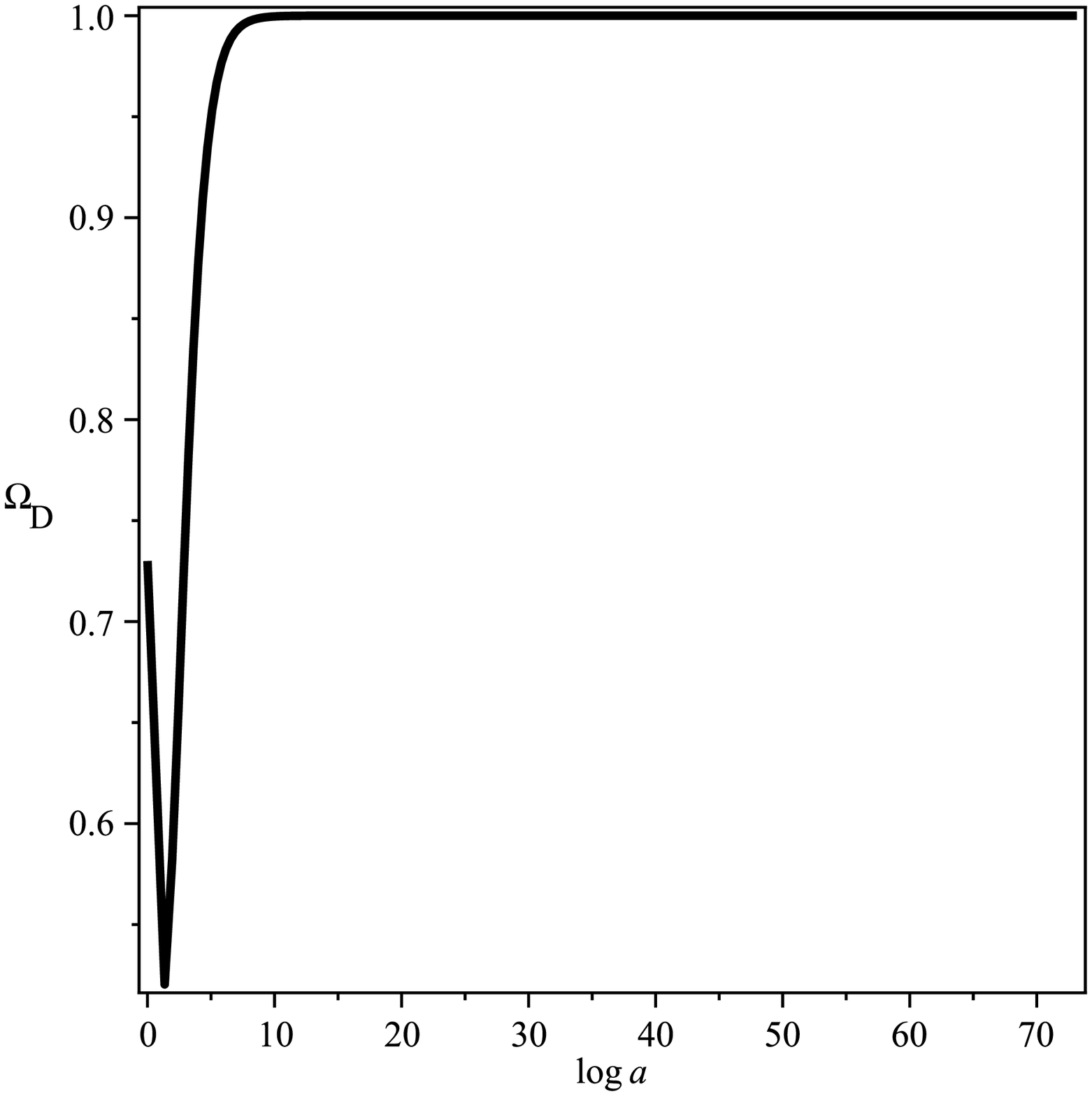} \\
\includegraphics[width=7cm]{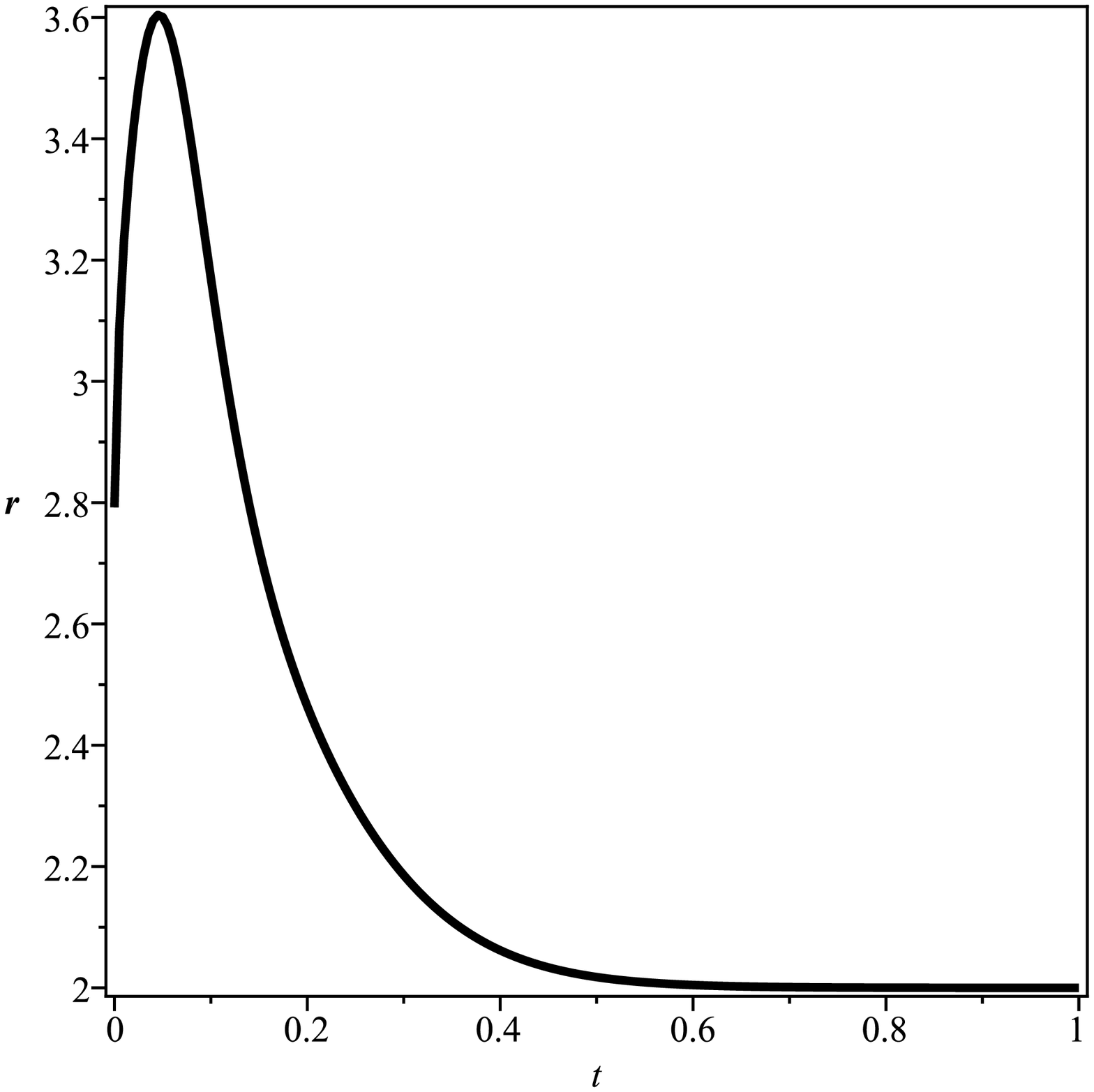}&
\includegraphics[width=7cm]{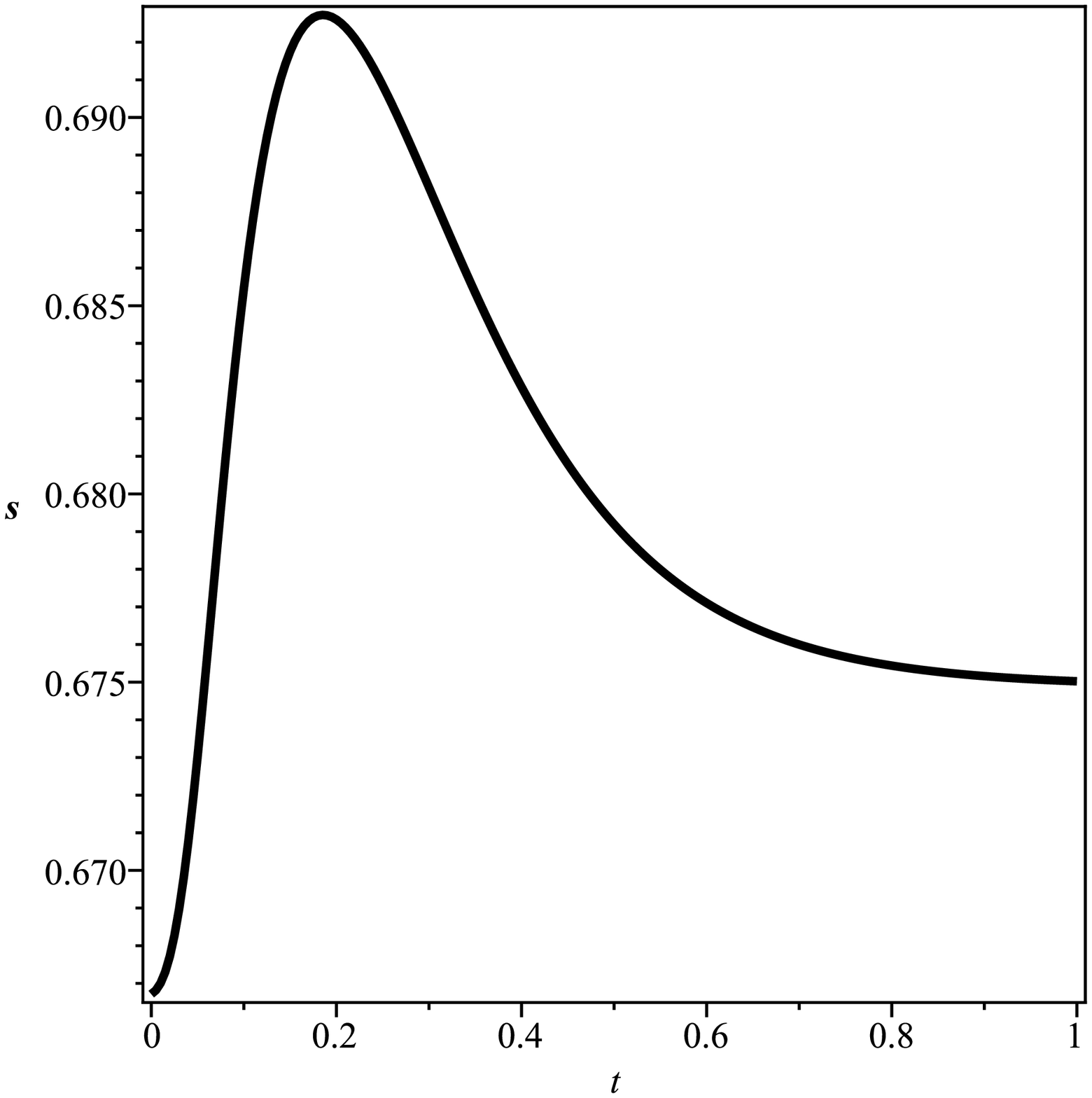} \\
\end{tabular}
\caption{(\textit{Top Left}) Time evolution of NADE density
parameter and logarithmic scale factor.    (\textit{Top Right})
Evolutionary behavior of NADE density parameter versus logarithmic
scale factor. (\textit{Bottom Left}) Time evolution of statefinder
parameter $r$. (\textit{Bottom Right}) Time evolution of statefinder
parameter $s$. While plotting, we adopted the following fixing of
free parameters: $a(0)=1$, $\dot{a}(0)=H_0=74$, $\Omega_D(0)=0.73$,
$n=0.5$, $\rho_1=0.187$. }
\end{figure*}

Figures shown with caption Fig.1 describe the evolution of cosmic
parameters of holographic dark energy. Here the top left and top
right figures, the evolution of dark energy density parameter and
evolution of scale factor. As is shown, the parameter
$\Omega_D\rightarrow1$ as $\ln a\rightarrow2$. Moreover the loop
quantum corrected $r$ and $s$ parameters are plotted in the bottom
two figures, which show that they asymptote to 2 and 0.66 in the
allowed range of time. Moreover the figures under the caption Fig.2
depict similar behavior quantitatively.

\section{Conclusion}

The statefinder parameters $r$ and $s$ have been calculated for a
Universe described by Loop quantum cosmology. Two dark energy
models, namely Holographic dark energy and New Agegraphic dark
energy have been considered and the statefinder parameters are
calculated for these models separately in loop quantum cosmology.
The trajectories in the r-t and s-t planes for these dark energy
models characterize their individual properties. Unfortunately, do
to complicated nature of statefinder parameters, we were unable to
plot r-s diagrams.

\end{document}